\documentclass[twocolumn,prl,amsmath,amssymb,amsfonts,groupedaddress,floatfix,showpacs]{revtex4}
\usepackage[dvips]{graphics}
\usepackage{subfigure}
\usepackage{graphicx}
\usepackage{amsmath}
\usepackage{color}

\newcommand{\Lg}{\textbf{L}}

\begin{document}

\title{Giant resistance change across the phase transition in spin crossover molecules}

\author{N. Baadji}
\affiliation{School of Physics and CRANN, Trinity College, Dublin 2, Ireland}
\author{S. Sanvito}
\affiliation{School of Physics and CRANN, Trinity College, Dublin 2, Ireland}

\begin{abstract}
The electronic origin of a large resistance change in nanoscale junctions incorporating spin crossover molecules is 
demonstrated theoretically by using a combination of density functional theory and the non-equilibrium Green's 
functions method for quantum transport. At the spin crossover phase transition there is a drastic change in the 
electronic gap between the frontier molecular orbitals. As a consequence, when the molecule is incorporated in a 
two terminal device, the current increases by up to four orders of magnitude in response to the spin change. This is 
equivalent to a magnetoresistance effect in excess of 3,000~\%. Since the typical phase transition critical temperature 
for spin crossover compounds can be extended to well above room temperature, spin crossover molecules appear as 
the ideal candidate for implementing spin devices at the molecular level.
\end{abstract}

\date{\today}

\maketitle


The last few years have witnessed several attempts at implementing spin devices at the single molecule level \cite{Spin01,Spin02,Spin03}.
These include spin-valves, where organic molecules are used as transport medium for spin-polarized electrons injected from magnetic 
metals \cite{Ralph,schmaus,Seneor}, and three terminal junctions incorporating magnetic molecules \cite{HvdZ}. The second 
strategy appears particularly intriguing as the electrical response of single molecule magnets depends sensitively on both 
their spin and charge configuration \cite{HvdZ,Das}. This sensitivity, combined with the possibility of manipulating the spin-state of a 
molecule by electrical means \cite{Loss,Nadjib}, makes spin-electronics based on magnetic molecules an attractive platform for both 
conventional and quantum logic. 

There are however two unfortunate drawbacks in constructing spin-devices based on molecular magnets. The first concerns with the typically poor 
structural stability of such molecules away from solution. For instance almost all the members of the Mn$_{12}$ family react on metallic surfaces, 
so that their structural and electronic integrity in uncertain once they are incorporated in a device \cite{Mn12break}. The second problem 
has to do with the magnetic anisotropy. In fact, although the anisotropy density per atom may be very large, the overall magnetic 
anisotropy of the molecule is rather small, since only a handful of atoms contribute to the magnetic moment. This means that 
at room temperature the spin quantization axis of the molecule fluctuates at a frequency much higher than the transport measurement 
time and the molecule appears paramagnetic to a steady state transport experiment. Importantly, while the first problem can be kept under 
control by engineering the end groups binding to the substrate \cite{Fe4}, the second one appears much more tough to overcome. 
In fact, quantum tunneling of the magnetization is always present so that the instability in the magnetization direction may persist down to low 
temperatures. It is then not surprising that the only demonstration to date of spin-valve effect originating from molecular magnets is limited to 
ultra-low temperatures \cite{ErSV,ErSV2}.

One possible way to overcome such a difficulty is to abandon the concept of spin-valve and to look at devices where the information is
contained in the actual spin magnitude and not in its direction. A natural choice for such alternative device concept is that of using molecules
displaying a magnetic bi-stability, i.e. molecules that can be found in two different spin states, both accessible with an external stimulus. 
This is the realm of spin-crossover compounds \cite{SCC}. Such materials are formed by 3$d$ transition metal ions in an octahedral 
surrounding, which display a spin transition from a low spin state (LS), usually the ground state, to high spin state (HS), usually a 
metastable one. Such a LS-HS transition can be triggered by temperature, pressure or light. It is accompanied by a modification 
of the geometrical structure, which alters the crystal field strength. The most extensively investigated spin crossover compounds 
contain Fe(II) and the transition is between a $^1$A$_{1g}$ LS state to a $^5$T$_{2g}$ HS one.  
  
To date electron transport experiments in spin crossover molecules at the single molecular level remain scarce, mostly because the effort in depositing 
such molecules on surfaces has intensified only recently. However, several deposition techniques are becoming available (e.g. 
Ref.~\cite{10.1063/1.3192355}) and a first demonstration of light-induced spin-crossover in vacuum-deposited thin films has been 
already reported \cite{Naggert}. Intriguingly the critical temperature for spin-crossover in such thin films is similar to that of 
the bulk, giving hope that room temperature device operation may be soon achievable. Also intriguing is the possibility to tune the 
spin crossover transition with an electric field, as demonstrated recently for valence tautomeric compounds \cite{andrea}. 

Importantly there are two demonstrations of the interplay between the spin-crossover transition and the electron transport properties 
of a device. Prins {\sl et al.} \cite{ADMA:ADMA201003821} measured a change in the current-voltage, $I$-$V$, curve of a 
Fe-(trz)$_3$ (trz=triazole) complex deposited between Au nano-gap electrodes, which correlates well with the spin-crossover 
transition. Furthermore, the three-terminal single molecule experiments of Meded {\sl et al.} \cite{PhysRevB.83.245415} showed 
low-energy features attributable to spin-crossover and controllable by gating. 

\begin{figure}[h]
\begin{center}
{\large{\bf(a)}}\includegraphics[scale=0.15,clip=true]{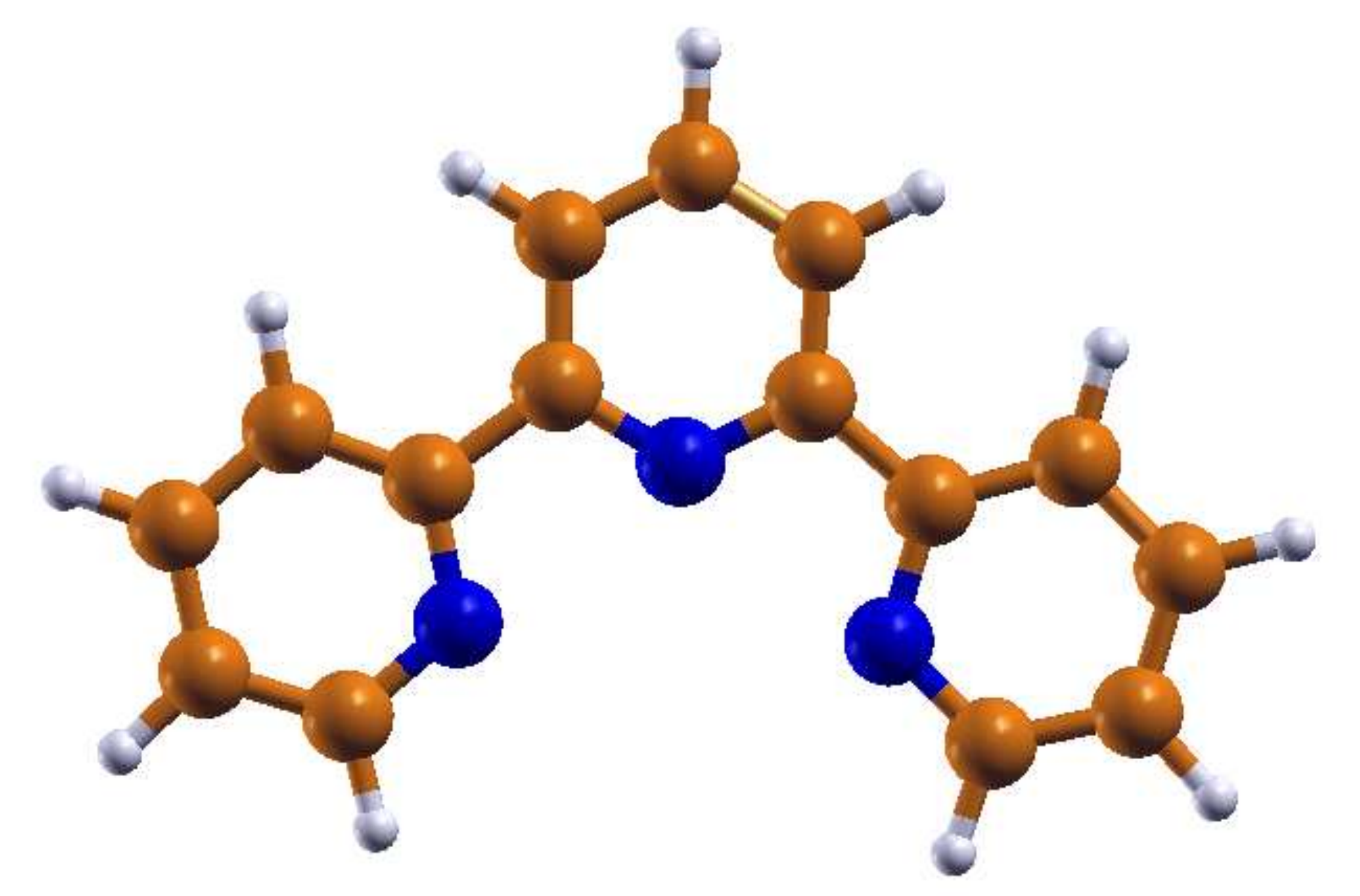}{\large{\bf(b)}}\includegraphics[scale=0.11,clip=true]{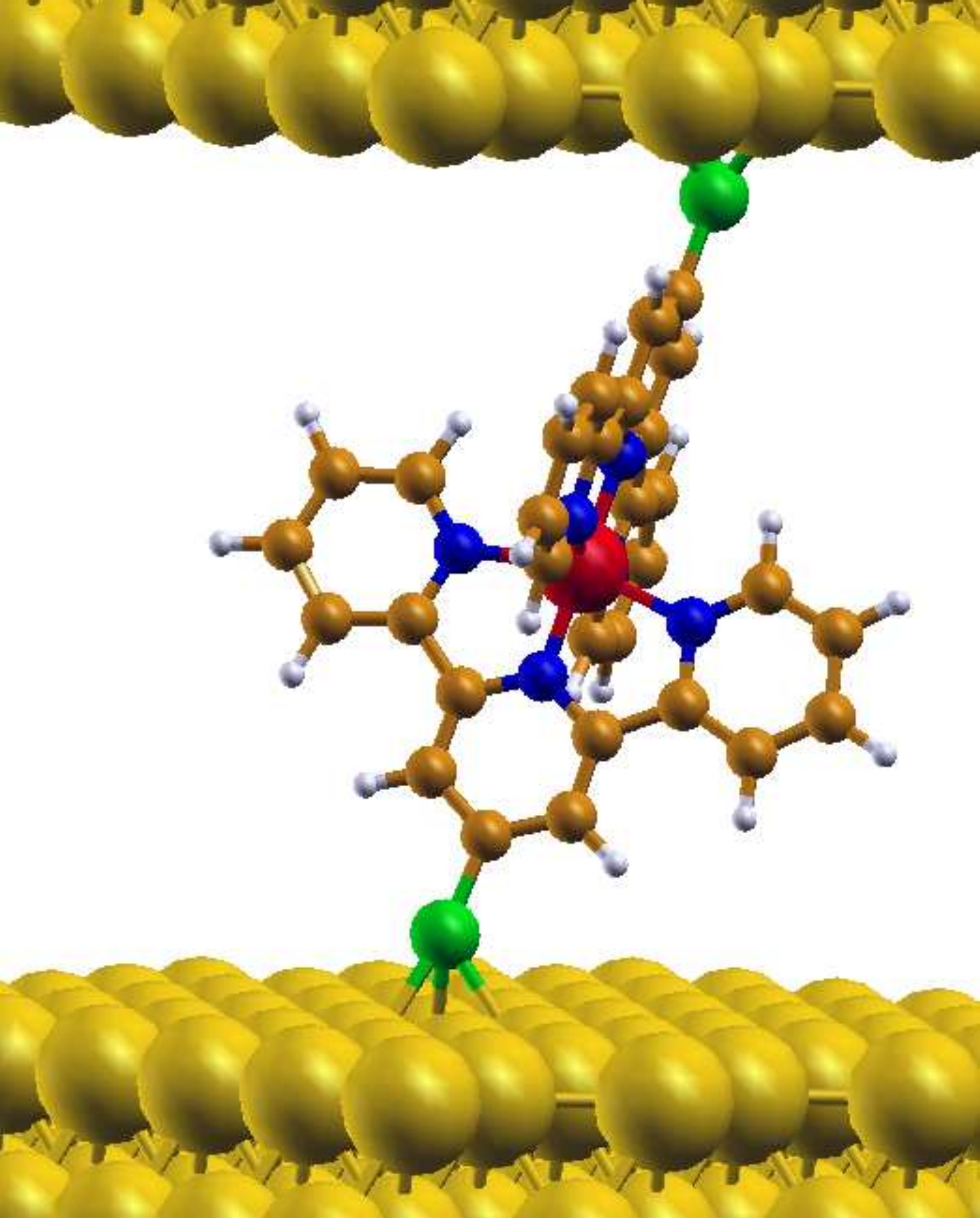}
\includegraphics[width=\columnwidth,keepaspectratio=true]{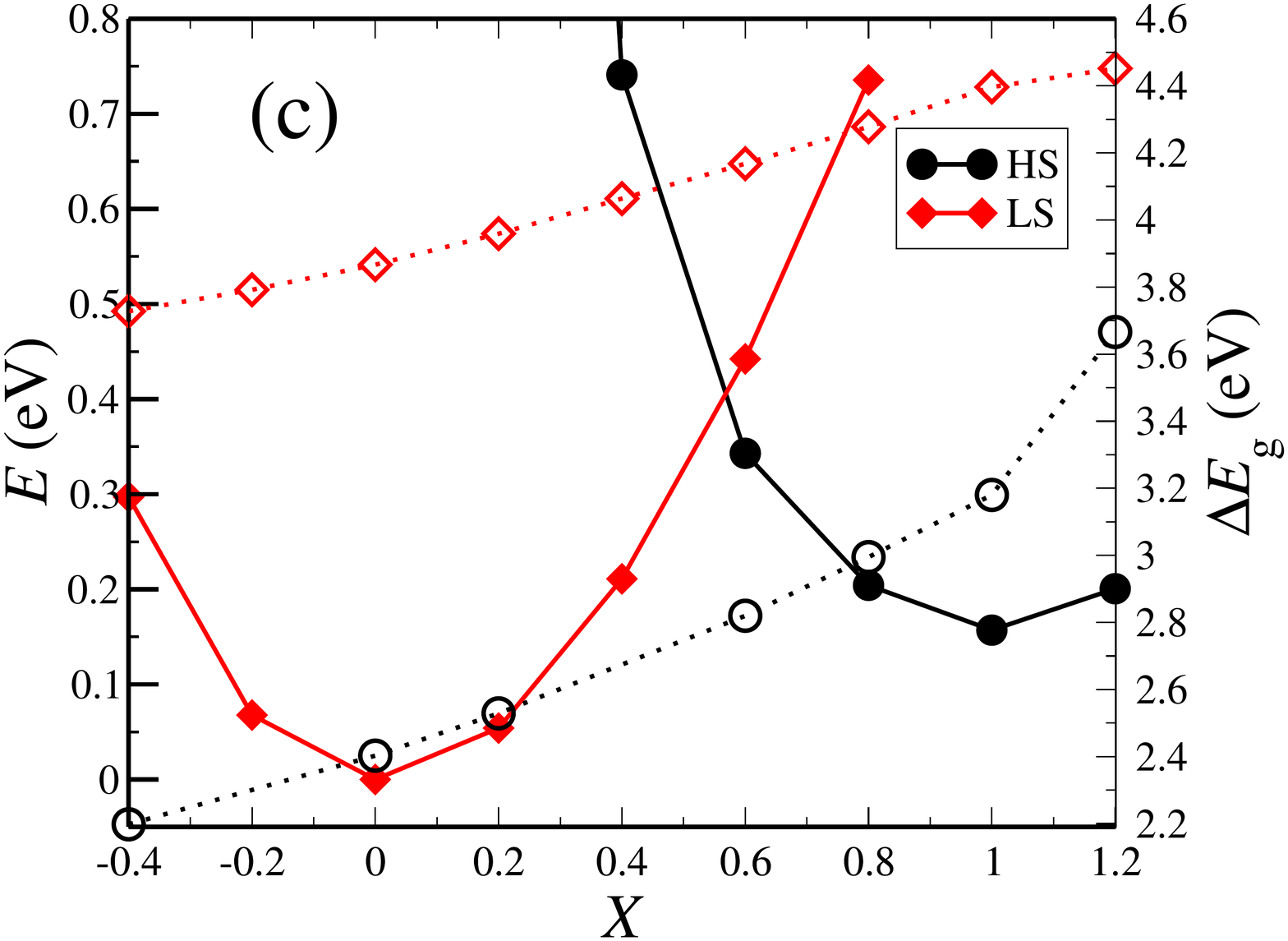}
\caption{(Color on line) Electronic properties of [Fe\Lg$_2$]$^{+2}$, \Lg~=~$2,2'$:$6,2''$-terpyridine, along the reaction coordinates, 
$X$. In (a) and (b) we show respectively the $2,2'$:$6,2''$-terpyridine group and the cell used for the transport calculations. Color 
code: C=orange, N=dark blue, H=white, S=green and Au=yellow. In panel (c) we present the total energy, $E$ (full symbols to read on 
the left-hand side scale), and the HOMO-LUMO gap, $\Delta E_\mathrm{g}$ (open symbols to read on the right-hand side scale), 
for both LS and HS along $X$. Calculations are performed with the B3LYP functional.} 
\label{Fig1}
\end{center}
\end{figure}

A crucial aspect of spin crossover compounds is that at the LS to HS phase transition there is a simultaneous electronic and structural change of 
the molecule. The electronic gap between the highest occupied molecular orbital (HOMO) and the lowest unoccupied molecular 
orbital (LUMO) varies accordingly to the different occupation of the 3$d$ multiplet and at the same time there is an out-breathing 
relaxation of the octahedral cage connected to the weakening of the crystal field. As such, in transport experiments it is difficult to 
separate effects arising from the electronic structure from those originating from the molecule geometry. In this letter we address 
such an issue with state of the art electron transport calculations based on density functional theory (DFT). In particular we show 
that the change in HOMO-LUMO gap at the phase transition may generate extremely large changes in the $I$-$V$ characteristic. 
This is demonstrated for the prototypical Fe(II) SCC [Fe\Lg$_2$]$^{+2}$, where \Lg~is a $2,2'$:$6,2''$-terpyridine group 
[see Fig.~\ref{Fig1}(a)]. Note that qualitatively similar results (not presented here) have been obtained for a second 
[Fe\Lg$_2$]$^{+2}$ compound where \Lg~=~$2, 6-$bis(pyrazol-1-yl), so that we expect our main conclusions to bare generality.

Calculating the electronic properties of spin crossover molecules with DFT is a rather difficult task as the relative energy separating the LS and HS 
state is very sensitive to the choice of exchange and correlation functional and to the atomic relaxation \cite{Casida} (note that 
the molecule atomic coordinates are known only for single crystals but not in vacuum or on a surface). We then use the following 
strategy. We first perform structural relaxation for the molecule in vacuum by using the B3LYP functional \cite{B3LYP}, one of the 
better performing for this problem. Then the results are compared with those obtained by using the generalized gradient 
approximation (GGA) and the local density approximation incorporating the self-interaction corrections (LDA-SIC) \cite{ASIC}. 
From this analysis we select the GGA functional to be used for the electron transport calculations, as this reproduces qualitatively 
the HOMO-LUMO gap obtained by B3LYP~\cite{gap}.

We start our analysis by discussing the electronic and geometric properties of the molecule in the gas phase. B3LYP calculations 
are performed with the NWCHEM package \cite{nwchem}. The basis set is 6-31++G$^{**}$ for H, C and N, while for Fe we use the 
Los Alamos double zeta with an effective core potential \cite{lanl2dz}. In Fig.~\ref{Fig1}(c) we present the total 
energy as a function of the reaction coordinates, $X$, interpolating between the LS ($X=0$) and the HS ($X=1$) geometry for 
the two different spin configurations. In the same figure we also show the dependence of the HOMO-LUMO gap, 
$\Delta E_\mathrm{g}$, as a function of $X$.

The figure clearly shows that the ground state is, as expected, low spin. As $X$ increases there is a transition to high spin for $X$$\sim$0.6 with 
the difference between the energy minima of the two spin configurations being about 200~meV. The HOMO-LUMO gap also changes as a function of 
$X$, increasing for both the spin states as one goes from the low to the high spin geometry ($X$ gets bigger), i.e. as the ligand field 
weakens. Most importantly at the relative energy minimum the gap, calculated with B3LYP, of the LS state is $3.87$~eV (for $X$=0), while 
that of the HS is only $3.18$~eV (for $X$=1). At the same two energy minima LDA-SIC~\cite{scaling} gives us HOMO-LUMO gaps of 
3.35~eV (LS) and 2.84~eV (HS), while GGA returns 1.81~eV (LS) and 0.33~eV (HS). Although different exchange and correlation functionals 
yield sensibly different gaps, in all cases this contracts significantly across the phase transition. Importantly both B3LYP and GGA give us 
spin down HOMO and LUMO levels in the HS state, while with LDA-SIC the HOMO is spin up and the LUMO is spin down. As such, 
we decide to use GGA instead of LDA-SIC for the transport calculations.

\begin{figure}
\begin{center}
\includegraphics[width=\columnwidth,keepaspectratio=true]{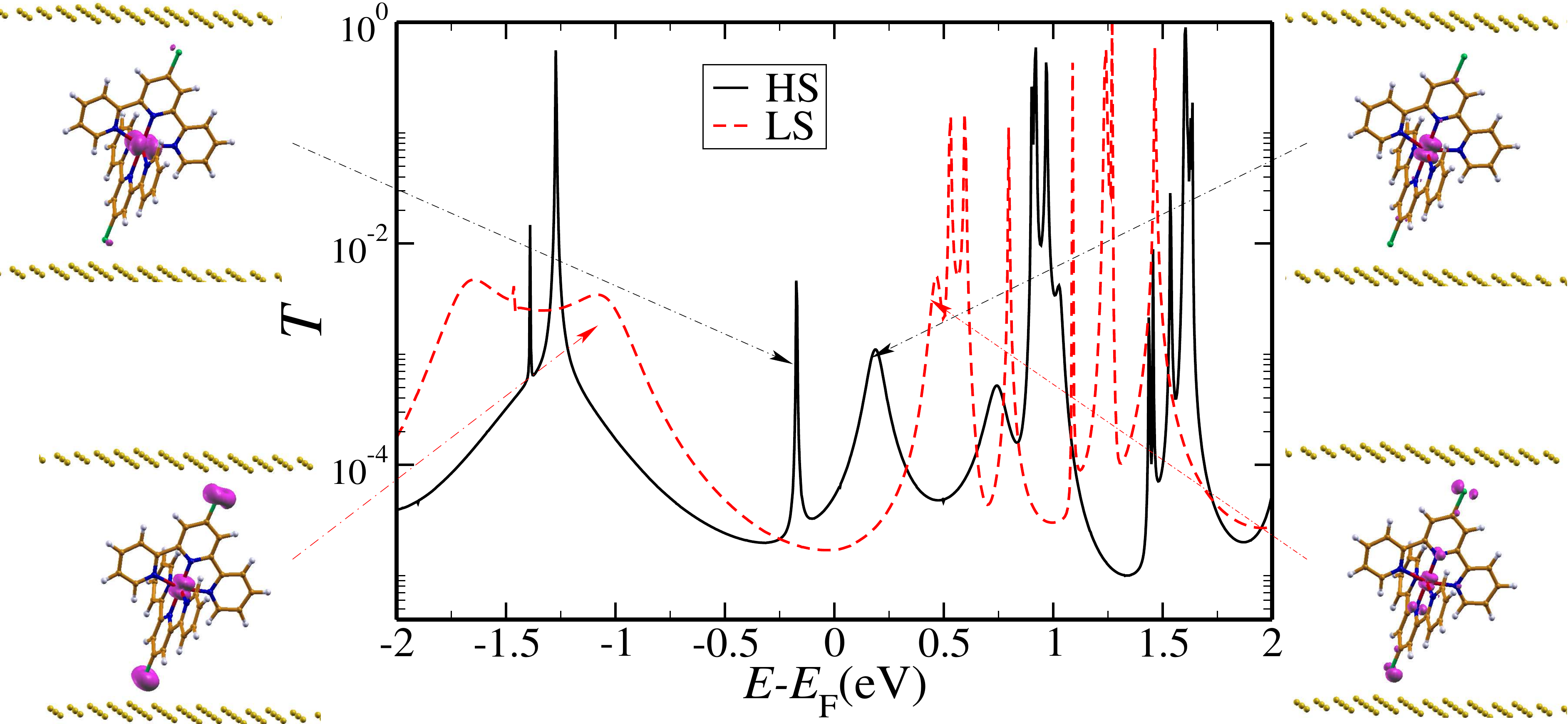}
\caption{(Color on line) Zero-bias transmission coefficient as a function energy, $T(E;V=0)$, for both the LS (red dashed line) and 
the HS (solid black line) configuration of the Fe\Lg$_2$-based junction and the molecular orbitals associated to their HOMO and 
LUMO like-orbital. Calculations are performed with the GGA functional.} 
\label{Fig2}
\end{center}
\end{figure}
We construct a two terminal device by attaching the Fe\Lg$_2$ molecule to a Au(111) surface via thiol group. In particular we consider 
bonding to the (111) hollow site and we optimize the Au-S distance for two different angles between the molecule and the normal 
to the surface, namely 0 and 30$^\mathrm{o}$. As the transport properties depend weakly on such tilting angle the results reported 
here refer only to the 30$^\mathrm{o}$ case, the bonding geometry presenting a lower energy. Transport calculations are performed 
with the {\it Smeagol} code \cite{Smeagol1,Smeagol2}, which implements the non-equilibrium Greens function method within 
DFT \cite{Siesta}. The wave-function is expanded over a numerical atomic orbital basis set and the core electrons are described with 
norm-conseving pseudopotentials including non-linear core corrections. The basis set has double zeta quality for C, N, H 
and Fe, while only single zeta is employed for Au. The simulation cell contains the molecule and 5 (111)-oriented Au 
layers with $8\times 7$ cross section. We use periodic boundary conditions in the direction orthogonal to the transport with a 
uniform $2\times 2$ $k$-point sampling and an equivalent mesh cutoff of 400~Ry. The charge density is integrated over 64 
energy points along the semi-circle, 64 energy points along the line in the complex plane and 64 poles are used for the Fermi 
distribution (the electronic temperature is 25~meV). During the finite-bias calculations we integrate the Green's function over the 
real axis on a 256 point mesh.

Figure \ref{Fig2} displays the total zero-bias transmission coefficient, $T(E;V=0)=\sum_\sigma^{\uparrow\downarrow} T^\sigma(E;V=0)$, 
as a function of energy for both the LS and HS configurations of the junction ($\sigma$ labels the spin, $\sigma=\uparrow, \downarrow$). 
The most striking feature is a 
radical change of the position of the various resonant peaks as the molecule transits from LS to HS. This is a direct consequence of the 
re-arrangement of the Fe $d$ shell density of state originating from the reduction of the ligand crystal field. In particular as the 
crystal field weakens in the HS state the energy width of $d$-shell reduces and the peaks become more dense. This effect is particularly 
dramatic around the Fermi level, $E_\mathrm{F}$, where the transport HOMO-LUMO gap shrinks. Recalling here the fact that in the 
linear response the conductance $G$ is simply given by $G=\frac{e^2}{h}T(E_\mathrm{F};V=0)$, we can conclude that at the 
phase transition there is a significant conductance enhancement. One can quantify such an effect by defining the spin-crossover 
magneto-resistance (SCMR) ratio as $R_\mathrm{SCMR}=({G_\mathrm{HS}-G_\mathrm{LS}})/{G_\mathrm{LS}}$. 
At zero-bias we find $R_\mathrm{SCMR}=200$~\%, a variation which is certainly measurable. Note that across the phase transition 
there is no change in the molecule valence charge, so that the effect reported here occurs at an energy scale lower than that needed 
to observe Coulomb blockade. 

\begin{figure}
\begin{center}
\includegraphics[width=7cm,keepaspectratio=true]{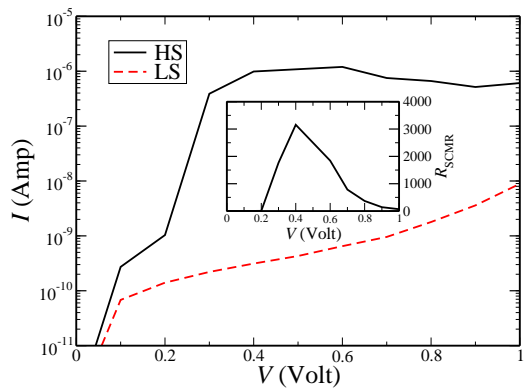}
\caption{(Color on line) $I$-$V$ curve for a Fe\Lg$_2$ molecule attached to gold electrodes in either the HS (black solid line) or 
the LS (red dashed line) state. The current is plotted on a logarithmic scale only for positive currents. The inset shows the 
spin-crossover magneto-resistance (SCMR) ratio, $R_\mathrm{SCMR}$. Calculations are performed with the GGA functional.} \label{figure-2}
\end{center}
\end{figure} 
Some more details are provided by the insets of Fig.~\ref{Fig2}, where we show the frontier molecular orbitals associated to 
the different peaks in $T(E;V=0)$. In general around $E_\mathrm{F}$ the orbitals responsible for the transmission have all
Fe-$d$ character and mix little with the ligands. In the LS configuration the HOMO and LUMO are non-spin-polarized
and have respectively $d_{z^2}$ and $d_\pi$ character. In contrast the HS state has a HOMO of $x^2-y^2$ symmetry, while
the LUMO is $d_\sigma$-like, but both the HOMO and the LUMO are spin down. Thus the gap reduction is the result of the 
orbital re-arrangement of the $d$ manifold associated to the spin crossover transition. In the LS state the gap is between the 
$t_{2g}$ triplet and $e_g$ doublet split by the octahedral crystal field. In the HS configuration the system undergoes Jahn-Teller 
distortion in the singly occupied spin down triplet and thus the gap reduces. Note that 
such a distortion, and thus the spin crossover transition, may be obstructed if the molecule is too rigidly bound to the electrodes.
As such for this effect to be detected one has to protect the spin active part of the molecule by appropriate chemical groups,
so that the Fe-N cage can deform freely

Then we look at the $I$-$V$ characteristics, Fig.~\ref{figure-2}. Clearly, the drastic gap reduction in the HS 
state makes its relative current considerably larger than that of the LS. At a voltage of about 0.5~Volt the two differ by 
more than three orders of magnitude. The LS current is essentially tunneling-like over the entire bias window investigated 
and it increases monotonically with bias. Such an increase is the result of the enlargement of the bias window, i.e. of the 
fact that a larger portion of the transmission spectrum contributes to the current. It is then the tail of the LUMO to 
contribute the most. The HS situation is different since voltages of the order of 0.4~Volt are enough for bringing both the 
HOMO and the LUMO transmission resonances within the bias window (note that orbital re-hybridization under bias 
drifts both the HOMO and LUMO close to each other hence decreasing the gap). As a consequence the current grows 
fast at low bias, when there is a transition from tunneling to resonant transport, and then saturates. Such a large difference 
in the $I$-$V$'s of the two spin states translates in a large $R_\mathrm{SCMR}$, which is presented as a function of bias 
in the inset of Fig.~\ref{figure-2}. Clearly $R_\mathrm{SCMR}$ as large as 3000\% appear possible.

\begin{figure}[t]
\begin{center}
\includegraphics[width=6cm,keepaspectratio=true]{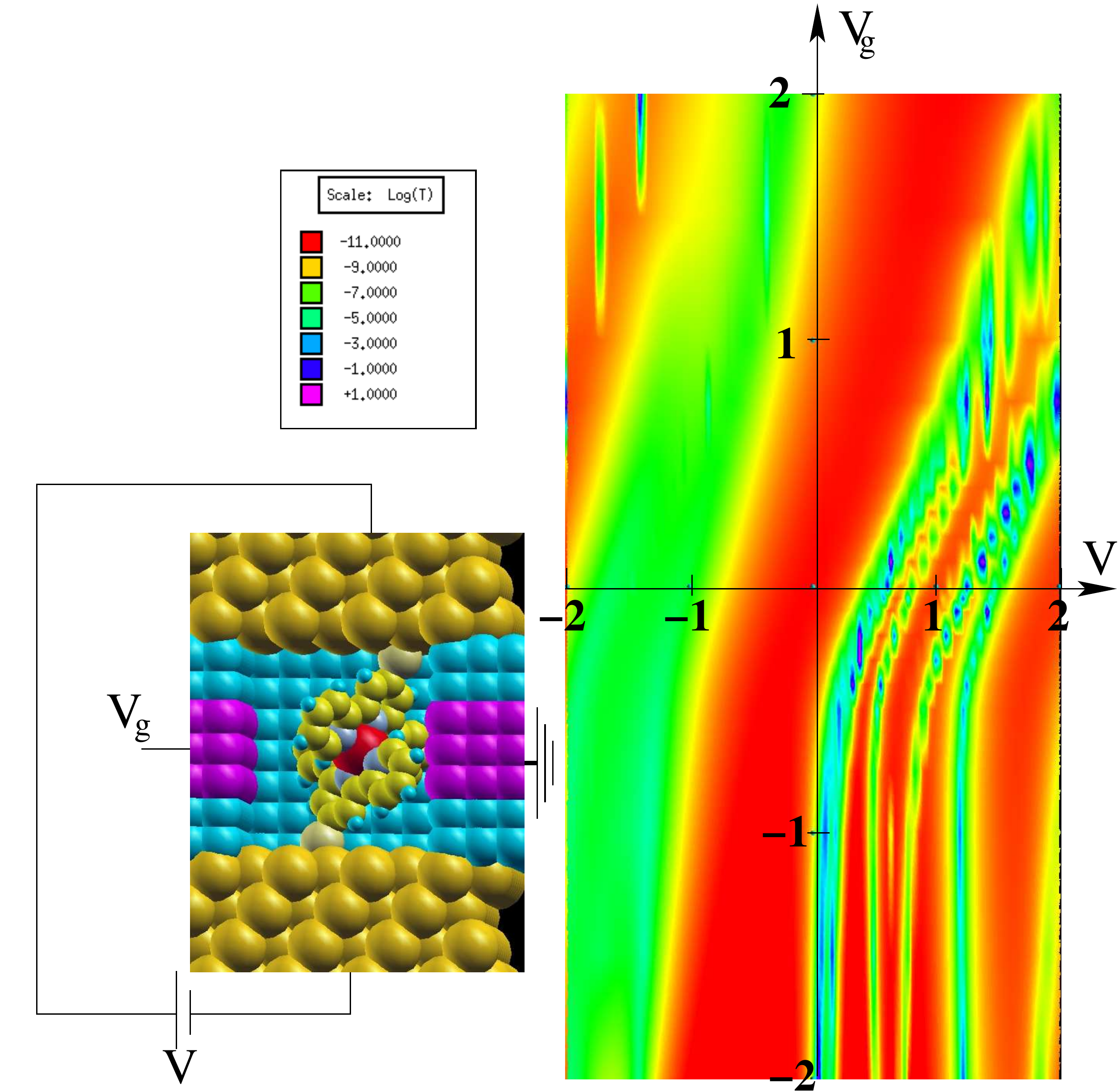}
\caption{(Color on line) Transmission coefficients as a function of energy for different values of the gate potential $V_\mathrm{g}$.
Here $T(E;V=0)$ is plotted as a color code on a logarithmic scale with red corresponding to low transmission (10$^{-11}$) and blue 
to high ($10^{-1}$). In the inset we show a schematic figure of the device investigated. The purple atoms represent the region in
space where the gate is applied. The color code for the atoms is the same as in Fig.~\ref{Fig1}(b).} 
\label{figure3}
\end{center}
\end{figure}
Finally, we look at the effect of a potential gate on the transport properties of the molecule. This is applied by shifting 
the on-site energies of the molecular orbitals \cite{Ivangate}. We note that for both states the transmission spectrum is characterized by a number of relatively sharp peaks spaced by at most 1.5~eV (most typically 
the spacing is $\sim$0.3~eV) and corresponding to states with substantial Fe $d$ character. These are expected to shift with
bias. As an example we consider the LS situation and in figure~\ref{figure3} we show $T(E;V=0)$ as a function of gate.
Here $T$ is plotted as a color code on a logarithmic scale with red corresponding to low transmission (10$^{-11}$) and blue 
to high ($10^{-1}$). Clearly the entire spectrum shifts with the gate potential until a state, either the HOMO or the LUMO, crosses
the Fermi level ($V=0$). At that point charge transfer takes place and the gate becomes less efficient. Crucially when one of the 
two frontier molecular orbitals is brought to $E_\mathrm{F}$ the zero-bias conductance increases dramatically. Since
the two spin states of the molecule have different frontier molecular orbitals, their position can be manipulated independently, i.e. 
they can be brought at $E_\mathrm{F}$ for different values of the gate. As such, an extremely large gate-modulated 
magnetoresistance is expected.

In conclusion we have demonstrated that the changes in electronic structure associated to the spin crossover phase transition, 
namely the reduction of the HOMO-LUMO gap, produce large changes in the transmission of a two-terminal molecular junction. 
At low bias we have predicted a spin crossover magnetoresistances of the order of 200\%, which can increase up to 3000\% at finite bias. 
The effect is related to the different nature of the frontier molecular orbitals of the different spin configurations and can be tuned
with a gate potential. Such an effect, whose experimental evidence has been already partially provided, can be exploited in
the fabrication of spin-transistors at the molecular level.  Intriguingly our results not only prove that spins can be manipulated in
an organic molecule, but also introduce a mechanism for molecular switching which has electronic origin and it is not associated
to a change in the molecule geometry \cite{SVDM}.

This work is supported by Science Foundation of Ireland under the NanoSci-E+ project ``Internet'' (08/ERA/I1759).


\begin{thebibliography}{99}
\bibitem{scaling} Here we set the scaling parameter $\alpha=0.7$.
\end{thebibliography}
\end{document}